\documentclass[12pt]{article}

\usepackage{graphicx}
\usepackage{pslatex}
\usepackage{amsfonts}

\title{Counting Independent Sets \\ and Kernels of Regular Graphs}
\date{}
\author{Adam B. Yedidia}

\begin{document}
\maketitle
\begin{abstract}
Chandrasekaran, Chertkov, Gamarnik, Shah, and Shin recently proved that the average number of independent sets of
random regular graphs of size $n$ and degree 3 approaches $w^n$ for large $n$, 
where $w \approx 1.54563$, consistent with the Bethe approximation.
They also made the surprising conjecture that the fluctuations of the logarithm of the number of independent sets were
only $O(1)$ as $n$ grew large, which would mean that the Bethe approximation is amazingly accurate for all 3-regular graphs.
Here, I provide numerical evidence supporting this conjecture obtained from {\em exact} counts of independent sets using
binary decision diagrams. I also provide numerical evidence that supports the novel conjectures that the
number of {\em kernels} of 3-regular graphs of size $n$ is given by $y^n$, where $y \approx 1.299$, and that the fluctuations
in the logarithm of the number of kernels is also only $O(1)$.
\end{abstract}

\thispagestyle{empty}
\newpage
\setcounter{page}{1}
\pagestyle{plain}
\section{Introduction}
In this paper I consider the problem of counting the number of independent sets and kernels of regular graphs. For the purposes of this paper, a {\em graph} will be defined as a collection of {\em vertices} and a collection of {\em edges} that connect pairs of vertices. {\em Simple} graphs are graphs with no more than one edge between any two vertices, and with no edges that connect a vertex to itself. The {\em degree} of a vertex is the number of edges connected to that vertex. A {\em k-regular} graph is a simple graph in which each vertex has degree $k$. An {\em independent set} is a set of vertices in a graph, no two of which are connected by an edge. A {\em kernel}, also called a ``maximal independent set,'' is an independent set such that adding any other vertex to the set forces the set to contain a pair of vertices connected by an edge.

Independent sets are closely related to ``hard sphere'' models that physicists use to model liquids and gases. In a hard sphere model, particles never overlap. Independent sets could therefore be seen as the legal positions of particles on a lattice, with no two particles being adjacent. In physics, the number of legal configurations of a hard sphere model is known as the ``partition function,'' and the logarithm of that number is known as the {\em entropy} of the model.

In a recent paper \cite{freakout}, Chandrasekaran et al. proved that the average number of independent sets for 3-regular graphs of size $n$ will approach $w^n$ as $n$ grows large, where $w \approx 1.54563$. This value was computed using the Bethe approximation from statistical physics \cite{Jed}. They also made a very surprising prediction about the fluctuations around this result. Ordinarily, the fluctuations between random samples of similar systems will grow as $\sqrt n$, where $n$ is the size of the system. Instead, Chandrasekaran et al. conjectured that the standard deviation of the logarithm of the number of independent sets of random regular graphs will not be $O(\sqrt n)$, as one might expect, but instead be $O(1)$! This implies that the Bethe approximation will always provide an amazingly accurate estimate for the entropy of independent sets for every randomly chosen 3-regular graph.

Chandrasekaran et al. proved that their surprising conjecture is true if the Shortest Cycle Cover Conjecture (SCCC) of Alon and Tarsi \cite{Alon} is true, but they offered no direct numerical evidence. This is most likely due to the difficulty of actually counting independent sets for large graphs. Counting independent sets, even for 3-regular graphs, is a \#P-hard problem \cite{sharp}, meaning that the time it takes to count independent sets will grow exponentially as the graph sizes grow. Well-suited, however, to counting solutions to combinatorial problems is the binary decision diagram (BDD), first introduced by Bryant \cite{Bryant}, and recently explicated by Knuth \cite{Knuth}. In my paper, I use BDDs to gather numerical evidence that convincingly confirms the conjecture of \cite{freakout}.

Chandrasekaran et al. did not make any predictions about kernels. We can still use the BDDs to gather evidence about kernels, however, and the evidence shows that kernels behave very similarly to independent sets. More precisely, I make the novel conjectures that the average number of kernels of 3-regular graphs grows as $y^n$, with $y \approx 1.299$, and that the fluctuations in the logarithm of that number are only $O(1)$ as $n$ grows large. 

\section{Independent Sets and Kernels}
In this section I will give a more detailed explanation of independent sets. Recall that an independent set is defined as a set of vertices in a graph, no two of which are connected by an edge.

\begin{figure}
  \centering
  \includegraphics[width=0.4\textwidth]{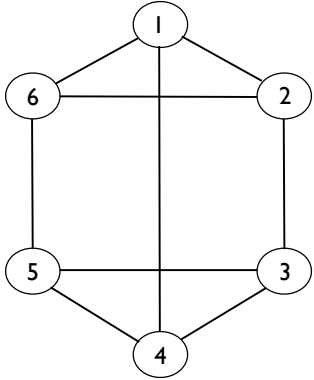}
  \caption {A regular 6-graph.}
\label{g6}
\end{figure}

Figure \ref{g6} shows a $3$-regular graph of six vertices (or ``6-graph''). The independent sets of this graph would be \{$\emptyset$, \{1\}, \{2\}, \{3\}, \{4\}, \{5\}, \{6\}, \{1, 3\}, \{1, 5\}, \{2, 4\}, \{2, 5\}, \{3, 6\}, \{4, 6\}\}, because those are the thirteen possible sets of vertices such that no two of them will be connected.

\begin{figure}
  \centering
  \includegraphics[width=0.35\textwidth]{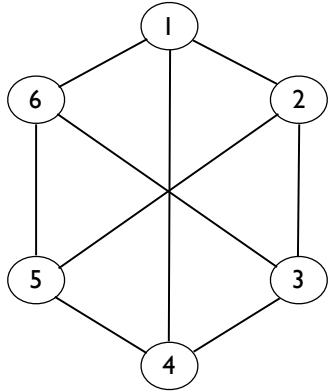}
  \caption {A different regular 6-graph.}
\label{g6_2}
\end{figure}

Change the figure to another 6-graph of degree 3, and the independent sets change as well, as in figure \ref{g6_2}. This graph's independent sets are \{$\emptyset$, \{1\}, \{2\}, \{3\}, \{4\}, \{5\}, \{6\}, \{1, 3\}, \{1, 5\}, \{2, 4\}, \{2, 6\}, \{3, 5\}, \{4, 6\}, \{1, 3, 5\}, \{2, 4, 6\}\}. There are fifteen configurations here; this shows that there can be variance in the number of independent sets even in graphs of identical size and degree. 

Recall now that a kernel is a ``maximal independent set'' or an independent set to which one cannot add a vertex without also adding an edge. Although there were 13 independent sets for the graph in figure \ref{g6}, there are only 6 kernels: they are \{\{1, 3\}, \{1, 5\}, \{2, 4\}, \{2, 5\}, \{3, 6\}, \{4, 6\}\}. The graph in figure \ref{g6_2} contains 15 independent sets, but there are only {\em two} kernels: \{\{1, 3, 5\}, \{2, 4, 6\}\}. This shows that more independent sets does not necessarily translate into more kernels, and also shows that there is perhaps more variance in the number of kernels than in the number of independent sets. However, this paper will show that as the number of vertices grows, the number of kernels and independent sets will actually behave very similarly.

Kernels and independent sets can also be thought of as binary functions of ones and zeroes. This is done by assigning each vertex of a graph a value of either 0 or 1. When checking if a possible configuration of 0's and 1's is an independent set or kernel, one considers the vertices included in the set to have value 1, and those that are excluded to have value 0. The binary function for an independent set (or kernel) has a value 1 if the configuration corresponds to an independent set (or kernel), and 0 otherwise.

\section{Binary Decision Diagrams}
Binary decision diagrams (BDDs) provide compact representations of binary functions \cite{Bryant}\cite{Knuth}; in our case the binary functions represent independent sets and kernels. 
Because BDDs are the source of all of the numerical evidence in this paper, it is essential that the paper contain an adequate explanation of them. 

Although graphs and BDDs look similar to
each other, they serve quite different purposes.
A BDD is composed of {\em nodes} and {\em links} between those nodes, only now the links ``flow'' in a particular direction and the relationship between the links and the nodes is more complicated than in a graph. Each node has a value, denoted {\tt V}, a {\tt LO} branch, which ``points'' to another node, and a {\tt HI} branch, which also points to another node. Each node's ({\tt LO, HI}) combination must be unique for it to be a true binary decision diagram, and at each node, ${\tt LO} \ne {\tt HI}$. At each node, {\tt V} describes the variable on which the decision depends. For example, in a graph of size $n$ as described above, it is often convenient to number the vertices of the graph $1, 2, 3,...,n$. So if we were to use a BDD to describe the binary function corresponding to independent sets, a node with ${\tt V}=x$ would depend on the vertex numbered $x$ in the graph. The {\tt LO} and {\tt HI} branches of this node would point to other nodes; the idea is that if the vertex numbered $x$ had a value of 1, one should take the {\tt HI} branch to the next node, and if it was equal to 0, one should take the {\tt LO} branch. These nodes will eventually point to two ``sinks,'' {\tt True} and {\tt False}. The sink one reaches by going down the tree will determine whether the path you have taken corresponds to
the binary function having a value of 1 ({\tt True}) or 0 ({\tt False}).

\begin{figure}
  \centering
  \includegraphics[width=.5\textwidth]{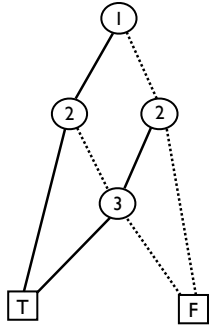}
  \caption {A BDD for the majority function.}
\label{BDD1}
\end{figure}

This idea is best explained with an example. Let us suppose we have three binary variables, $x_1$, $x_2$, and $x_3$, and let us suppose our binary function is the ``majority function'' which has value 1 if and only if two or more of the three variables have value 1. The BDD for this problem would look like figure \ref{BDD1}.

As we look at this BDD, we consider first the top node. In BDDs, a solid line denotes the {\tt HI} path and a dotted line the {\tt LO} path. Let us assume that $x_1=1$ and therefore we take the {\tt HI} path, to the leftmost 2 node. Here, we see that if we take the {\tt HI} path again, we go directly to the {\tt True} sink, without even considering 3. This is because once we know that both $x_1$ and $x_2$ equal 1, we already know that the majority function equals 1--it doesn't matter what $x_3$ is. In fact, it would be incorrect to add the redundant extra node: we stated earlier that no node in a BDD can have ${\tt LO}={\tt HI}$, and both of the extra node's branches would point to {\tt True}.

Let us go back to the leftmost 2 node. If we choose the {\tt LO} path, then that means $x_1=1$ and $x_2=0$. That means that for the majority function to equal 1, $x_3$ must equal 1. That is why $x_3$'s {\tt HI} branch points to {\tt True} and its {\tt LO} branch points to {\tt False}: this final variable decides the value 
of the majority function.

\begin{figure}
  \centering
  \includegraphics[width=0.8\textwidth]{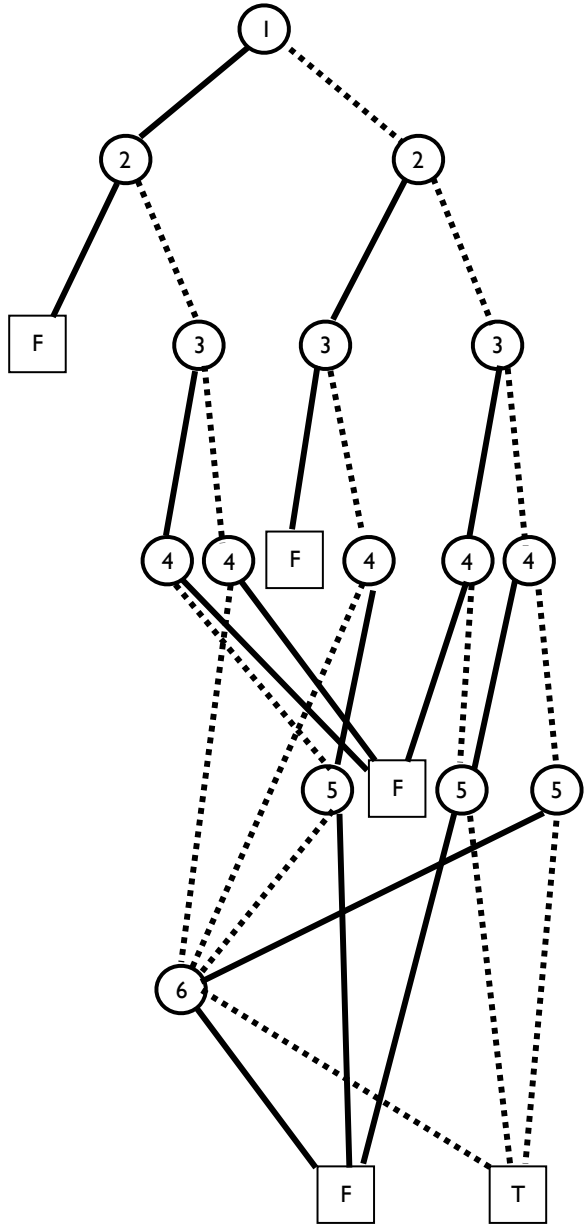}
  \caption {A BDD representing the independent sets of a 3-regular 6-graph.}
\label{BDD2}
\end{figure}

Now we consider a more complex BDD. Figure \ref{BDD2} is the BDD for the independent sets of the first graph we looked at. Extra {\tt False} sinks have been added for a clearer picture.

This BDD is complicated, but one can still recognize some patterns. For example, one can only reach {\tt False} sinks by taking a {\tt HI} branch--this makes sense, since removing a vertex from an independent set always yields another independent set. 

The main use of BDDs such as this in this paper, however, is not to be read by humans, but to be read by computer. We exploit the fact that there exist algorithms to systematically construct the BDD for the independent sets and kernels of a graph \cite{Knuth}. Moreover, given a BDD, it is straightforward to {\em exactly} count the number of solutions of the binary function it represents, which in our case corresponds to the number of independent sets (or kernels). The counting algorithm \cite{Knuth} works as follows, where $s$ is the total number of nodes in the BDD, counting the {\tt True} and {\tt False}
nodes as one node each, and $v_k$, $l_k$, and $h_k$ are {\tt V}, {\tt LO}, and {\tt HI} for the $k$th node.
\begin{itemize}
\item Step 1: [Loop over $k$]. Set $c_0 \leftarrow 0$, $c_1 \leftarrow 1$, and do Step 2 for $k=2,3,...,s-1$. Then return the answer $2^{v_{s-1}-1}c_{s-1}$.
	
\item Step 2: [Compute $c_k$]. Set $l \leftarrow l_k$, $h \leftarrow h_k$, $c_k \leftarrow 2^{v_l-v_k-1}c_l+2^{v_h-v_k-1}c_h$.
\end{itemize}
Using this algorithm, it is possible to quickly and efficiently count solutions to independent sets of reasonably small graphs. This is what I did for graphs of degree 3 and sizes 6 to 40 (even numbers only, because it is impossible to have a graph of odd degree and odd size). My data convincingly confirm the conjecture of Chandrasekaran et al. for independent sets, and lead to similar conjectures for kernels. 

\section{Numerical evidence}\label{results}

To create the data presented in this section, I generated 1000 random 3-regular simple graphs of each even size between 6 and 40. It is easy to do this using an algorithm that randomly adds edges between vertices that still have fewer than three edges, and that have not previously been connected. For each graph I then created a BDD by automatically generating, for that graph, appropriate input for D. E. Knuth's BDD creation program \cite{bdd14} written in his ``BDD language.'' The above counting algorithm was used to exactly count the number of solutions of each BDD. 

For both independent sets and kernels, the BDDs were created using a boolean function that was a large {\tt AND} function 
of a collection of local
functions. For the independent set case, each local function 
required that the variables corresponding to the two vertices on an edge were not both 1. For the
kernel case, the local functions required for each vertex variable that
if it was 
was 1, all its neighbors' vertex variables were 0, and if it was 0, at least one of its neighbors' vertex
variables was 1. This last condition corresponds to the requirement that one cannot add a vertex to a maximal
independent set and have it remain an independent set.

Knuth's program records the number of memory accesses it makes as it creates a BDD. Memory accesses in modern computers
dominate the running time, so they serve as a good proxy for computational complexity. I found that the average number of
memory accesses to create a BDD for the independent sets of a 3-regular graph of size $n$ grew roughly as
$400 \times 1.28^n$. Because of the exponential growth in the complexity, BDD's, like any other algorithm for exact counting of
independent sets and kernels, are limited to relatively small $n$.

\subsection{Independent sets}

Chandrasekaran et al. prove that at $n$ grows large, the average number of independent sets of a 3-regular $n$-graph will approach $w^n$, where $w=z^{-3/2}(2-z)^{-1/2}$ and $z$ is a root of the equation $z^3+z-1=0$, giving $w \approx 1.545634155$. 

\begin{figure}
	\centering
\begin{tabular}{c|c|c|c}
	$n$ & $w^n$ & mean & $w_{est} = e^{\frac{\ln mean}{n}}$\\ \hline
	6 & 13.635 & 13.464 & 1.5423952668\\ \hline
	8 & 32.573 & 31.815 & 1.54109350802 \\ \hline
	10 & 77.815 & 75.777 & 1.54153624619\\ \hline
	12 & 185.9005 & 181.494 & 1.54254741637\\ \hline
	14 & 444.1134 & 434.487 & 1.54321669622\\ \hline
	16 & 1060.980 & 1041.904 & 1.54388245415 \\ \hline
	18 & 2534.665 & 2485.237 & 1.54394400334\\ \hline
	20 & 6055.279 & 5930.353 & 1.5440239311\\ \hline
	22 & 14465.97 & 14191.04 & 1.54428663307\\ \hline
	24 & 34558.98 & 33960.44 & 1.54450939167\\ \hline
	26 & 82560.89 & 81049.27 & 1.54453602897\\ \hline
	28 & 197236.7 & 193795.5 & 1.54466285137\\ \hline
	30 & 471195.6 & 462317.9 & 1.54465451307\\ \hline
	32 & 1125679 & 1106305 & 1.54479583718\\ \hline
	34 & 2689230 & 2639377 & 1.54478373281\\ \hline
	36 & 6424531 & 6313624 & 1.54488668558\\ \hline
	38 & 15348108 & 15109601 & 1.54499725062\\ \hline
	40 & 36666398 & 36075768 & 1.54500677979\\ \hline
\end{tabular}
\caption{Numerical results for the number of independent sets in random regular graphs, compared with the Bethe
approximation estimate of \cite{freakout}, which says that the mean should approach $w^n$, with $w\approx 1.54563$, 
for large $n$.}
\label{table1}
\end{figure}

Numerically, we can estimate $w$ for any $n$ as $w_{est} = \exp \left( \frac{\ln mean}{n} \right)$, where $mean$ is the
numerically determined mean of the number of counts. Figure \ref{table1} presents a table that shows that even using graphs of size $n=40$ or less, we could numerically estimate $w$ accurately to three significant figures if we
did not  know its exact value. Note that the estimate
of $w$
seems to be approaching its ultimate exact value from below.

Chandrasekaran et al.'s conjecture about fluctuations was precisely stated as follows in their Theorem 11 \cite{freakout}: 

``{\sl Let $G$ be chosen uniformly at random among all 3-regular graphs with $n$ vertices. Assuming SCCC is true, there exists a function $f:(0,1) \rightarrow \mathbb{R}^+$, so that $|\ln Z-\ln Z_B| \le f(\epsilon)$ with probability $1-\epsilon$, where $\frac{1}{n}\ln Z_B \approx \ln 1.545$.}''

Here $Z$ is the number of independent sets, $Z_B$ is the Bethe approximation to that number, and
``SCCC'' is the ``Shortest Cycle Cover Conjecture,'' due to Alon and Tarsi \cite{Alon} \cite{freakout},
which states ``{\sl Given a bridgeless graph $G$ with $m$ edges, all of its edges can be covered by a collection of cycles with the sum of their lengths being at most $7m/5=1.4m$.}''

Chandrasekaran et al.'s Theorem 11 means that for any 
probability $1-\epsilon$, the fluctuations in the number of independent sets of regular 3-graphs will not be more than $f(\epsilon)$. Since $f(\epsilon)$ does not depend on the size
of the graph, that means that the fluctuations are $O(1)$. The data that follows in this section is meant to test this
claim (which depends on the unproven SCCC), by numerically finding the function $f(\epsilon)$.

\begin{figure}%[htbp]
  \centering
  \includegraphics[width=0.6\textwidth]{a.1}
  \caption {Estimate of $f(\epsilon)$ for independent sets of 3-regular 6-graphs.}
\label{reg6}
\end{figure}

\begin{figure}%[htbp]
  \centering
  \includegraphics[width=0.6\textwidth]{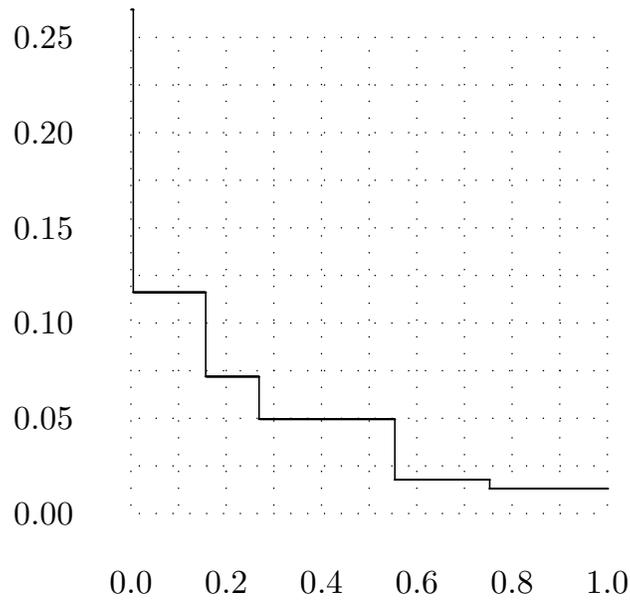}
  \caption {Estimate of $f(\epsilon)$ for independent sets of 3-regular 8-graphs.}
\label{reg8}
\end{figure}

The approach I take is to plot a numerical estimate $f(\epsilon)$, by computing the
difference $|\ln Z-\ln Z_B|$ for each graph, and
then finding the probability $1-\epsilon$ that the difference has a particular value for each $n$.
Consider for example the plot shown in figure \ref{reg6}, which is for the $n=6$ case. 
This graph only contains two values on the vertical axis: one at approximately 0.0954, the other at approximately 0.0476. This is because, as was mentioned before, there can only be 13 or 15 independent sets of a regular 6-graph of degree 3, and those are the differences one finds (in the logarithm of the number) with respect to the Bethe approximation of approximately 13.635. As the number of variables becomes larger, however, the number of ``levels'' in the graph will also increase. For example, figure \ref{reg8} shows the estimate of $f(\epsilon)$ for regular 8-graphs.

\begin{figure}%[htbp]
  \centering
  \includegraphics[width=0.6\textwidth]{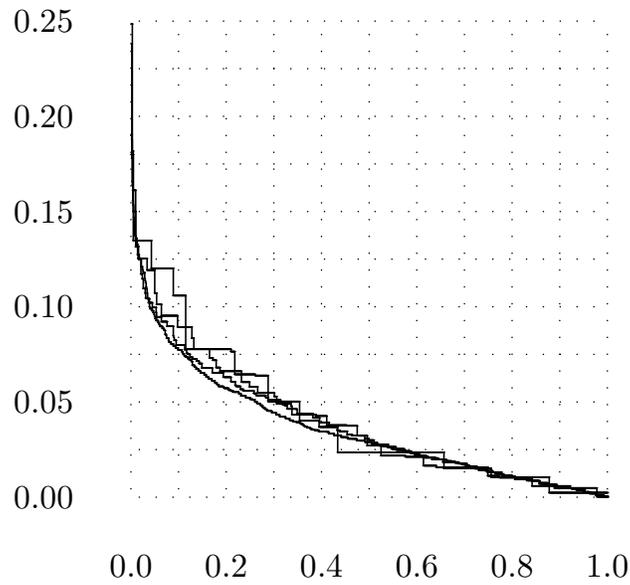}
  \caption {Estimates of $f(\epsilon)$ for independent sets of 3-regular $n$-graphs, with $n=10,12,14,16$.}
\label{reg10-16}
\end{figure}

The value of $f(0)$, which is the largest difference found between the true logarithm of the number
of independent sets and the Bethe approximation, increased 
from approximately .0954 to approximately .2646. This trend, however, will not continue, substantiating the prediction of Chandrasekaran et al. The largest difference $f(0)$ actually drops between 8 and 10, and will stabilize as the sizes get larger. So will the estimate for $f(\epsilon)$, for general $\epsilon$, which stops looking like a series of step function and start taking on a smoother shape. Figure \ref{reg10-16} is actually four plots, for 3-regular graphs of sizes 10, 12, 14, and 16, superposed.

\begin{figure}%[htbp]
  \centering
  \includegraphics[width=0.6\textwidth]{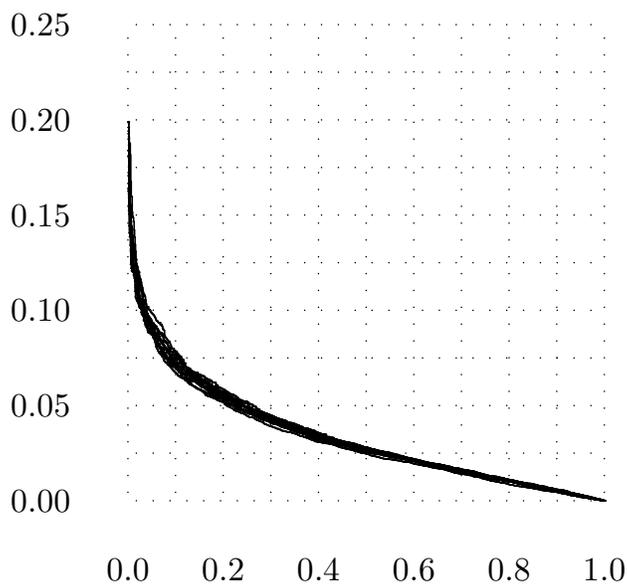}
  \caption {Estimates of $f(\epsilon)$ for independent sets of 
3-regular $n$-graphs, with $n$ taking all even values between $18$
and $40$, inclusive.}
\label{reg18-40}
\end{figure}

Figure \ref{reg18-40} shows the data for the remaining 12 plots, all superposed. Their sizes are comprised of the even numbers between 18 and 40, inclusive.
It is difficult to believe that the figure is twelve different sets of data. Not only do the fluctuations not grow beyond 
some upper limit (which is all that is necessary for the Chandrasekaran et al.'s conjecture to be true); they hardly change at all! This means we can confidently estimate the expected size of the fluctuations
in the entropy from our numerical data.

\begin{figure}
  \centering
  \includegraphics[width=0.4\textwidth]{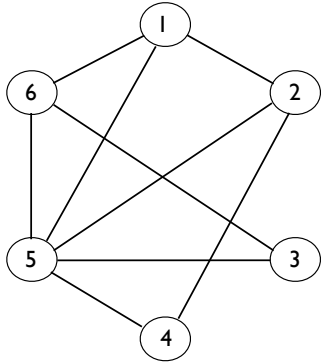}
  \caption {A 6-graph with average degree 3, that is not 3-regular.}
\label{r6}
\end{figure}

To emphasize how unusual the behavior shown in figure \ref{reg18-40} is, I
will present a similar set of data for random graphs that are selected to have the same size and the
same average degree of 3, but are not
necessarily regular. An example of a non-regular random 6-graph with average degree 3 is shown in figure \ref{r6}.
This graph's independent sets are \{$\emptyset$, \{1\}, \{2\}, \{3\}, \{4\}, \{5\}, \{6\}, \{1, 3\}, \{1, 4\}, \{2, 3\}, \{2, 6\}, \{3, 4\}, \{4, 6\}, \{1, 3, 4\}\}, so it has 14 independent sets. This graph was not previously possible, of course, because vertices 3 and 4 only have two edges, and vertex 5 has five.

\begin{figure}%[htbp]
  \centering
  \includegraphics[width=0.6\textwidth]{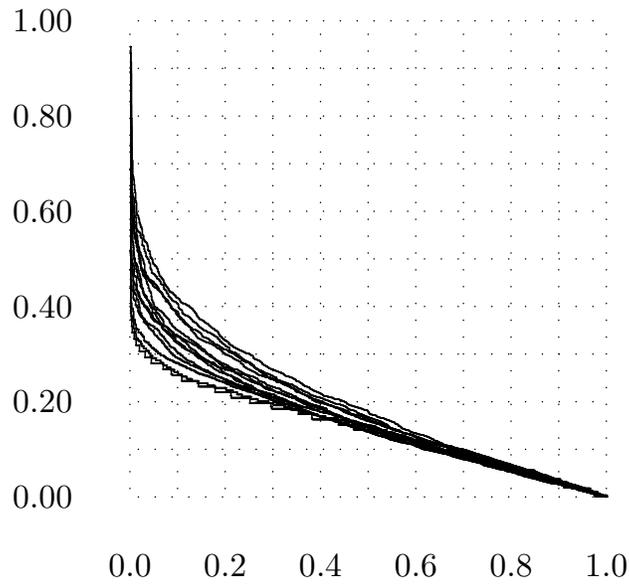}
  \caption {Estimates of $f(\epsilon)$ for independent sets of 
$n$-graphs with average degree 3, with $n$ taking all even values between $10$
and $36$, inclusive. The fluctuations consistently grow as $n$ increases.}
\label{rand10-36}
\end{figure}

For the class of $n$-graphs with average degree 3, the average number of independent sets is somewhat larger than it is
for random 3-regular graphs. I find that the average number of independent sets 
grows as $x^n$, where $x \approx 1.594$. For the 
fluctuations, one can measure a function $f(\epsilon)$ defined by $f(\epsilon) = |\ln Z - n \ln x|$.
Figure \ref{rand10-36} shows the estimated $f(\epsilon)$, obtained in the same way as the regular graph data, for the random graphs with average degree 3, and for $n$ between 10 and 36. Here, the fluctuations are clearly increasing with $n$,
as one would expect.

%\begin{figure}%[htbp]
%  \centering
%  \includegraphics[width=0.6\textwidth]{a.5}
%  \caption {Data for random 6-graphs.}
%\label{rand6}
%\end{figure}

%It looks similar to the data for the regular graphs of size 6 and degree 3, except now there is a third level, caused by the possibility of the graph in figure \ref{r6}.

%\begin{figure}%[htbp]
%  \centering
%  \includegraphics[width=0.6\textwidth]{a.6}
%  \caption {Data for random 8-graphs.}
%\label{rand8}
%\end{figure}

%If the error seems at first sight to have decreased, that is because of the changed dimensions of the graph. The highest error is .4868--nearly twice as much as the highest ever error for the regular graphs! The same dimensions will be used for figure \ref{rand10-36}, which superposes all of the data for graphs of even sizes between 10 and 36, inclusive.

%Figure \ref{rand6-36} shows a superposition of all the plots of 

%The error seems convincingly to be growing at a constant rate, confirming that only {\em regular} graphs have the property that they grow without increasing the error.

\subsection{Kernels}

\begin{figure}
	\centering
\begin{tabular}{c|c|c}
	$n$ & mean & $y_{est} = e^{\frac{\ln mean}{n}}$ \\ \hline
	8 & 7.941 & 1.29564015538 \\ \hline
	10 & 14.437 & 1.30601358862 \\ \hline
	12 & 23.420 & 1.30056553464 \\ \hline
	14 & 39.822 & 1.30105155128 \\ \hline
	16 & 66.855 & 1.30038175746 \\ \hline
	18 & 112.229 & 1.29985445627 \\ \hline
	20 & 189.283 & 1.29973729397 \\ \hline
	22 & 321.368 & 1.30003386341 \\ \hline
	24 & 540.124 & 1.29973224901 \\ \hline
	26 & 904.901 & 1.29931791247 \\ \hline
	28 & 1516.237 & 1.29896911345 \\ \hline
	30 & 2581.067 & 1.29935147154 \\ \hline
	32 & 4333.530 & 1.29912609539 \\ \hline
	34 & 7308.847 & 1.29910009294 \\ \hline
	36 & 12285.019 & 1.29895400448 \\ \hline
	38 & 20694.544 & 1.29889831749 \\ \hline
	40 & 34996.192 & 1.29897481351 \\ \hline
\end{tabular}
\caption{Table showing the mean number of kernels, averaged over 1000 3-regular $n$-graphs for each value of $n$,
and the numerical estimate for $y$, where the average is given by $y^n$.}
\label{table3}
\end{figure}

Next we look at the data for kernels.  First, I find that the average number of kernels for 3-regular $n$-graphs is approximately equal to
$y^n$, with $y \approx 1.299$. This value of $y$ can be read off from the table presented in figure \ref{table3}. Notice
that $y$ actually seems to reach its ultimate value more quickly than $w$ did.

\begin{figure}%[htbp]
  \centering
  \includegraphics[width=0.6\textwidth]{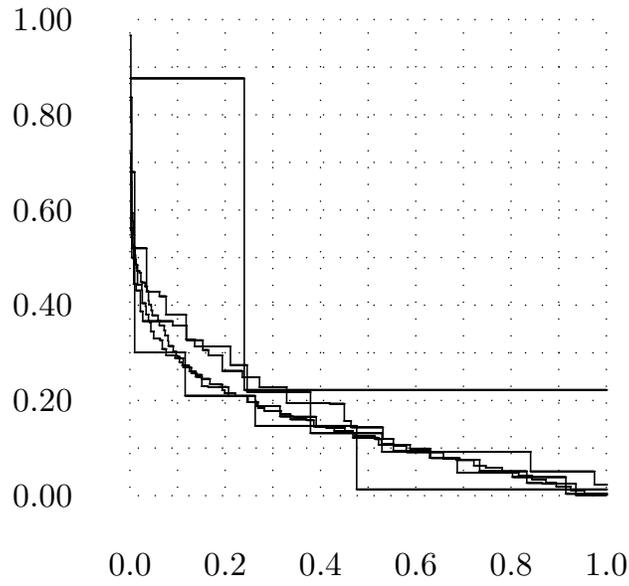}
  \caption {Estimates of $f(\epsilon)$ for kernels of 3-regular $n$-graphs, with $n=6,8,10,12,14,16$.}
\label{kern8-16}
\end{figure}

\begin{figure}%[htbp]
  \centering
  \includegraphics[width=0.6\textwidth]{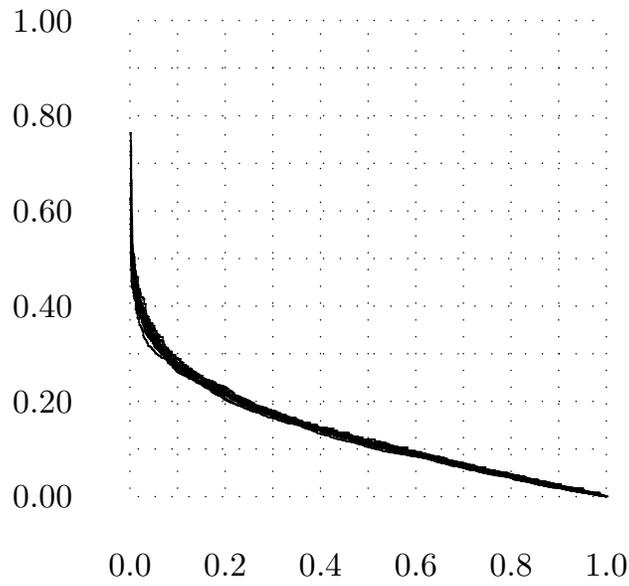}
  \caption {Estimates of $f(\epsilon)$ for kernels of 3-regular $n$-graphs, with $n$ taking all
even values from 18 to 40, inclusive.}
\label{kern18-40}
\end{figure}

For the plots of the fluctuations, one can estimate a function $f(\epsilon)$ analogous to the function for the
independent sets 
using $f(\epsilon) = |\ln Z - n \ln y|$, with $y=1.299$. With that in mind, figure \ref{kern8-16} shows the estimated
function $f(\epsilon)$ for kernels of 3-regular $n$-graphs, with $n$ ranging over even numbers from 6 to 16 inclusive,
while figure \ref{kern18-40} shows $f(\epsilon)$ for $n$ ranging from 18 to 40.

The estimated function $f(\epsilon)$ which measures fluctuations for kernels looks similar to that for independent sets, albeit with larger fluctuations.
This is not so surprising, if one recalls that regular 6-graphs have either 2 or 6 kernels, while they have 13 or 15 independent sets. 
Comparing figure \ref{kern18-40} for kernels with figure \ref{reg18-40} for independent sets, we see that the
fluctuations in the entropy are nearly four times as large for kernels as independent sets. It is also clear, however,
that numerically, 
kernels and independent sets share the essential property that their fluctuations do not grow as the graph size
increases. I thus conjecture that the fluctuations in the logarithm of the number of kernels in 3-regular $n$-graphs will
only be $O(1)$ as $n$ grows large.

The strong similarity between the numerical 
results for kernels and independent sets suggests that a Bethe approximation \cite{Jed} could give
a highly accurate result for the number of kernels. However, performing 
such a calculation turns out to be considerably more intricate for
the case of kernels than it was for independent sets, because the binary function representing configurations
that are kernels
is the {\tt AND} of local functions of vertex variables 
that involve
a vertex and all its neighbors (e.g. four variables in the case of 3-regular graphs), 
while for independent sets the local functions only involve simple
pairs of vertices. I hope to report on the results of such a calculation in the near future.

\section*{Acknowledgements}
I thank Jonathan Yedidia for his encouragement and advice.

\newpage

%\subsection{Computational Complexity}
%
%This section explores the amount of time and computation it took to gather the information presented in the rest of the %section. Figure \ref{table4} shows the number of memory accesses it took, averaged over 100 trials, to generate a BDD for %the independent sets of a graph of size $n$.

%\begin{figure}
%	\centering
%\begin{tabular}{c|c}
%	$n$ & $e^{\frac{\ln mean}{n}}$ \\ \hline
%	6 & 3082 \\ \hline
%	8 & 3115 \\ \hline
%	10 & 4900 \\ \hline
%	12 & 7636 \\ \hline
%	14 & 12243 \\ \hline
%	16 & 20269 \\ \hline
%	18 & 32689 \\ \hline
%	20 & 51825 \\ \hline
%	22 & 91696 \\ \hline
%	24 & 150101 \\ \hline
%	26 & 239844 \\ \hline
%	28 & 413666 \\ \hline
%	30 & 671363 \\ \hline
%	32 & 1102352 \\ \hline
%	34 & 1837845 \\ \hline
%\end{tabular}
%\caption{Table 4.}
%\label{table4}
%\end{figure}

\end{document}